\documentclass[10pt,doublecolumn,twoside]{IEEEtran}

\usepackage{etoolbox}
\newtoggle{doublecolumn}
\newtoggle{calculateFigures}

\toggletrue{doublecolumn}
\togglefalse{calculateFigures}
\usepackage{etex}

\usepackage[english]{babel}
\usepackage{lipsum}
\usepackage{amsbsy,amsmath,amsfonts,amssymb,amsthm}
\usepackage{mathtools}
\usepackage{textcomp} 
\usepackage{relsize}

\usepackage{bm,cite,cases,pstricks,times,url,verbatim} 
\usepackage[noend]{algpseudocode}
\usepackage{booktabs}
\usepackage{tabularx,array,dcolumn,multirow}

\usepackage{yfonts}

\usepackage{dutchcal} 

\usepackage{soul} 

\usepackage{blkarray,bigdelim} 

\allowdisplaybreaks[4]

\usepackage{centernot} 

\newcommand{\subparagraph}{}

\iftoggle{doublecolumn}{
    \newtheorem{thm}{Theorem}
    \newtheorem{fact}{Fact}
    \newtheorem{lemma}{Lemma}
    \newtheorem{definition}{Definition}
    \newtheorem{conj}{Conjecture}
    \newtheorem{propos}{Proposition}
    \newtheorem{corol}{Corollary}
    \newtheorem{ass}{Assumption}
    \newtheorem{example}{Example}
    \newtheorem{remark}{Remark}
    \newtheorem{note}{Note}
    \newtheorem{obs}{Observation}
}{

    \newtheoremstyle{exampstyle}
      {0} 
      {0} 
      {\itshape} 
      {} 
      {\bfseries} 
      {.} 
      {.5em} 
      {} 

    \theoremstyle{exampstyle} \newtheorem{thm}{Theorem}
    \theoremstyle{exampstyle} 
    \theoremstyle{exampstyle} 
    \theoremstyle{exampstyle} 
    \theoremstyle{exampstyle} 
    \theoremstyle{exampstyle} \newtheorem{propos}{Proposition}
    \theoremstyle{exampstyle} 
    \theoremstyle{exampstyle} 
    \theoremstyle{exampstyle} 
    \theoremstyle{exampstyle} 
    \theoremstyle{exampstyle} 
    \theoremstyle{exampstyle} 
}

\makeatletter
\newcommand{\pushright}[1]{\ifmeasuring@#1\else\omit\hfill$\displaystyle#1$\fi\ignorespaces}
\newcommand{\pushleft}[1]{\ifmeasuring@#1\else\omit$\displaystyle#1$\hfill\fi\ignorespaces}

\begingroup
\catcode`\#=11
\gdef\noautorotate{-dAutoRotatePages#/None}
\endgroup

\usepackage{graphicx,xcolor,float,dblfloatfix}
\usepackage{psfrag}

\usepackage{caption}
\usepackage{subcaption}

\newcommand{\subalign}[1]{%
  \vcenter{%
    \Let@ \restore@math@cr \default@tag
    \baselineskip\fontdimen10 \scriptfont\tw@
    \advance\baselineskip\fontdimen12 \scriptfont\tw@
    \lineskip\thr@@\fontdimen8 \scriptfont\thr@@
    \lineskiplimit\lineskip
    \ialign{\hfil$\m@th\textstyle##$&$\m@th\textstyle{}##$\crcr
      #1\crcr
    }%
  }
}

\iftoggle{calculateFigures}{
    \usepackage[crop=off,runs=2,pspdf=\noautorotate]{auto-pst-pdf}
}{
    \usepackage[off]{auto-pst-pdf}
}

\usepackage{hyperref}

\graphicspath{%
{./Figures/}
}

\usepackage{caption}
\begin{document}

\author{\IEEEauthorblockN{Alessandro~Biason and~Michele~Zorzi}\\
\IEEEauthorblockA{\{biasonal,zorzi\}@dei.unipd.it\\
Department of Information Engineering, University of Padova - via Gradenigo
6b, 35131 Padova, Italy
}%
}

\title{Energy Harvesting Communication System with SOC-Dependent Energy Storage Losses}

\maketitle



\begin{abstract}
The popularity of Energy Harvesting Devices (EHDs) has grown in the past few years, thanks to their capability of prolonging the network lifetime. In reality, EHDs are affected by several inefficiencies, \emph{e.g.}, energy leakage, battery degradation or storage losses. In this work we consider an energy harvesting transmitter with storage inefficiencies. In particular, we assume that when new energy has to be stored in the battery, part of this is wasted and the losses depend upon the current state of charge of the device. This is a practical realistic assumption, \emph{e.g.}, for a capacitor, that changes the structure of the optimal transmission policy. We analyze the throughput maximization problem with a dynamic programming approach and prove that, given the battery status and the channel gain, the optimal transmission policy is deterministic. We derive numerical results for the energy losses in a capacitor and show the presence of a \emph{loop effect} that degrades the system performance if the optimal policy is not considered.
\end{abstract}

\section{Introduction}

One of the most interesting techniques to prolong network lifetime is Energy Harvesting (EH), that allows the nodes of a Wireless Sensor Network (WSN) to refill their batteries with a renewable energy source, \emph{e.g.}, sunlight, vibrations, wind, etc. Differently from traditional networks, where the key aspects are energy conservation and efficiency, in an EH-WSN one of the most important aspects is \emph{energy management}, \emph{i.e.}, in order to improve the communication performance, the Energy Harvesting Devices (EHDs) have to correctly exploit the available resources.

A lot of research has focused on ideal devices~\cite{Lei2009,Sharma2010,Ozel2010,Ozel2012b}. In reality, EHDs have several inefficiencies, \emph{e.g.}, energy leakage~\cite{Devillers2012}, battery degradation~\cite{Michelusi2013c}, imperfect knowledge of the State Of Charge (SOC)~\cite{Michelusi2014} or storage losses~\cite{Tutuncuoglu2015}. In this work we focus on this last problem. In~\cite{Tutuncuoglu2015}, Tutuncuoglu \emph{et al.} assumed that the energy losses can be described with a fixed constant in $[0,1]$. 
Our work, instead, assumes that the storage losses \emph{depend upon the current SOC}, which is a realistic assumption, \emph{e.g.}, for a capacitor~\cite{Gorlatova2013}. Typically, when the battery is depleted or almost fully charged, it is not possible to store a lot of energy (high losses), whereas, when the battery is half-charged, almost all the harvested energy can be stored (low losses). With these considerations, the structure of the traditional optimal transmission policy changes. In this work we find the new optimal policy using a Markov Decision Process approach. Analytically, we prove the threshold structure of the optimal policy, \emph{i.e.}, for every SOC and channel gain, the optimal policy always uses the same transmission power. We present numerical results in order to describe the properties of the optimal policy. In particular, we notice a \emph{loop effect} (when the battery is almost empty and does not manage to be refilled), that degrades the system performance when the optimal policy is not used. Also, we show that the traditional optimal policy (obtained ignoring storage losses) is strictly sub-optimal in this context. Finally, we describe how the battery size influences the system throughput.

The paper is organized as follows. Section~\ref{sec:system_model} defines the system model and the optimization problem. In Section~\ref{sec:OP} we present the structure of the optimal policy. Section~\ref{sec:numerical_evaluation} presents our numerical results. Finally, Section~\ref{sec:conclusions} concludes the paper.

\section{System Model}\label{sec:system_model}

We consider an Energy Harvesting Device (EHD) with energy losses in the battery storage process. Time is divided in slots of equal length. In slot $k$, EHD harvests $B_k$ Joule of energy according to some energy arrival statistic. The harvested energy is divided in two parts: the first fraction, $I_k B_k$, is directly sent into the channel, whereas the second one, $(1-I_k)B_k$, recharges the device battery. $I_k \in [0,1]$ is the \emph{power splitting parameter} and is dynamically chosen by the device in each slot $k$. The battery has a finite size of $E_{\rm max}\ \mbox{J}$ and the energy stored in slot $k$ can be exploited only in a later slot. In an ideal situation, all the energy $(1-I_k) B_k$ is stored in the battery. Here instead, we consider a \emph{real battery} where only a fraction of $(1-I_k)B_k$ can be stored, while the rest is wasted due to the battery inefficiencies, a behavior that depends upon the current State Of Charge (SOC) $E_k$ of the battery. For example, in a capacitor~\cite{Gorlatova2013}, when SOC is very high or very low, it is not possible to store a lot of energy (very \emph{high losses}), whereas, when SOC is approximately half of the maximum storable energy, then almost all the energy can be stored (very \emph{low losses}). This inefficiency can be described by a function $s(y)$, where $y$ is the current SOC. An example, that was proposed in \cite{Gorlatova2013} and will be used as a baseline in this work, is~
\begin{align}
    s(y) &= 1-\frac{(y-E_{\rm max}/2)^2}{\beta_{\rm n.l.} (E_{\rm max}/2)^2}, \label{eq:s_x_y} 
\end{align}

\noindent which is a symmetric function with respect to $E_{\rm max}/2$. If the harvested power is $\omega$ (assumed constant in a time slot), then a power of $\omega\cdot s(y)\ \mbox{W}$ can be converted to energy and stored in the battery. When $y = 0$ or $y = E_{\rm max}$, then only $\omega(1-1/\beta_{\rm n.l.})\ \mbox{W}$ can be converted and $\omega / \beta_{\rm n.l.}\ \mbox{W}$ are wasted. Ideally, EHD wants to operate at $E_k = E_{\rm max}/2$ because $s(E_{\rm max}/2) = 1$. $\beta_{\rm n.l.}$ is a technology parameter strictly greater than one, \emph{e.g.}, $\beta_{\rm n.l.} = 1.05$. When $\beta_{\rm n.l.} \rightarrow \infty$, we have an ideal battery. We remark that the wasted power cannot be used in any other way, \emph{e.g.}, sent directly into the channel, because it is necessary to recharge the battery. In this paper, we use~\eqref{eq:s_x_y}, proposed in~\cite{Gorlatova2013}, as an approximate model to describe the energy losses in a capacitor. However, our analytical results do not depend upon the particular structure of $s(y)$, and can be used with other models as well.

The energy drawn from the battery and used for transmission in slot $k$ is $P_k \leq E_k$. Transmitting with power $\rho$ provides a reward $g(\rho,h)$, where $h$ is a parameter related to the channel status. The \emph{reward function} $g(\rho,h)$ is increasing and concave in $\rho$. An example, that we use as baseline in our numerical evaluation, is $g(\rho,h) = \ln(1+\Lambda h \rho)$, \emph{i.e.}, $g(\rho,h)$ represents the transmission rate ($\Lambda$ is an SNR scaling factor). Thus, in slot $k$ the reward is $g(P_k+B_k I_k,H_k)$ ($H_k$ represents the channel gain in slot $k$, i.i.d. over time and constant in a single slot), \emph{i.e.}, it is given by the sum of the energy drawn from the battery, $P_k$, and the part of the harvested energy directly sent into the channel, $B_k I_k$.
A time slot is structured as follows. At the beginning of the slot, EHD decides $P_k$ and $I_k$ according to its current energy level $E_k$ and to the channel gain $H_k$. During the slot, the device transmits and harvests energy with constant power. We assume we have only a causal energy arrival information, thus $B_k$ is unknown at the beginning of slot $k$.

Equation~\eqref{eq:s_x_y} is an instantaneous expression. To convert it to an amount of energy, we have to consider the battery evolution during a slot. Suppose that at the beginning of the slot the energy level is $y_0$. After a time $t$, the energy level is~
\begin{align*}
    y_t = y_0 + \int_0^t x s(y_\tau)\mbox{d}\tau
\end{align*}

Note that $y_T$ depends upon 1) the energy stored in the slot, namely $x$ (that is given by the combination of harvested and transmission power), 2) the slot length $T$ and 3) the initial battery level $y_0$. We explicitly write $y_T$ as $y_T = y_T(x,y_0)$. In the next we assume a fixed slot length, thus we interchangeably use the notions of energy or power.

The battery evolution is described by the following relation:~
\begin{align}
    E_{k+1} = \min\{y_T((1-I_k)B_k-P_k,E_k), E_{\rm max} \} \label{eq:batt_update}
\end{align}

\noindent Note that also $I_k$ influences the battery evolution. We include $P_k$ in the first argument of $y_T(\cdot,\cdot)$ because, during the slot evolution, the battery level changes also according to the transmission power ($(1-I_k)B_k-P_k$ represents the stored energy in slot $k$). The $\min$ is used because \emph{battery overflow} may occur.

In this paper we consider energy as measured in discrete energy quanta, and model the system with a discrete Markov Chain (MC). The MC state $(e,h) \in \mathcal{E} \times \mathcal{H} \triangleq  \{0,\ldots,e_{\rm max}\} \times \{0,\ldots,h_{\rm max}\}$ corresponds to $e$ energy quanta stored in the battery and to a channel gain $h$. When $e_{\rm max}$ energy quanta are stored, the battery is fully charged at $E_{\rm max}\ \mbox{J}$. The harvested energy can assume values in $\mathcal{B} = \{0,\ldots,b_{\rm max}\}$. 
We also apply a \emph{Round} operation to $y_T(\cdot,\cdot)$ and without loss of generality, we require $Round(y_T(b_{\rm max},e')) \geq 1$, where $e'$ is the energy state with the smallest $y_T(b_{\rm max},e')$. If this were not possible, then, when EHD is in state $e'$, its battery level could not increase. Note that the problem can always be reformulated by changing the notion of energy quanta in order to realize the previous condition. When the number of states is sufficiently large, these assumptions are not critical and the discrete model converges to the continuous one.

\subsection{Optimization} \label{sec:optimization}

A policy $\mu$, for every MC state, determines the probability of performing a certain action, \emph{i.e.}, $\mu$ is a set of random variable pairs $\Theta_{e,h}$ and $\Phi_{e,h}$ (one for every MC state) with joint probability density functions $f(\rho,\iota ; e,h)$. Each function maps an action pair $(\rho,\iota) \in \{0,\ldots,e\}\times [0,1]$ (transmission power and power splitting parameter) to a value in $\mathbb{R}^+$.

If $\mu$ induces a Markov Chain with at most one recurrent class, the long-term average reward does not depend upon the initial state $\mathbf{S}_0 = (E_0,H_0)$. In this case, the long-term average reward $G_\mu$ can be expressed as~
\begin{align}
    G_{\mu} &= \liminf_{K \rightarrow \infty} \frac{1}{K} \mathbb{E}_{B_k,H_k,P_k,I_k}\left[\sum_{k = 0}^{K-1} g(P_k+B_k I_k, H_k) | \mathbf{S}_0 \right] \nonumber \\
    &= \sum_{e \in \mathcal{E}} \pi_\mu(e) j_\mu(e), \label{eq:G_mu}
\end{align}

\noindent where $\pi_\mu(e)$ is the steady state probability of being in state $e$ and $j_\mu(e) \triangleq \mathbb{E}_{B,H}[ \mathbb{E}_{\Theta_{e,H},\Phi_{e,H}} [g(\Theta_{e,H} + B \Phi_{e,H},H)]]$. 
$\mathbf{S}_k = (E_k,H_k)$ is the \emph{state of the system} in slot $k$. $B$ and $H$ are the random variables of the energy arrivals and of the channel gain (i.i.d. over time) with means $\bar{b}$ and $\bar{h}$. Their pmfs are $p_B(b)$, $b \in \mathcal{B}$ and $p_H(h)$, $h \in \mathcal{H}$, respectively (note that we assume a discrete channel).

In this work we want to find the Optimal Policy (OP) $\mu^\star$ such that~
\begin{align}
    \mu^\star = \arg \max_{\mu} G_{\mu} \label{eq:mu_star}
\end{align}

Therefore, the problem is formulated using the theory of Markov Decision Processes, and can be solved with standard stochastic optimization techniques, \emph{e.g.}, the Policy Iteration Algorithm (PIA)~\cite{Bertsekas2005}. In Proposition~\ref{propos:admissible} we show that the optimal policy induces a MC with one recurrent class.

\section{Optimal Policy} \label{sec:OP}

The aim of this section is to prove a fundamental property of OP.
\begin{thm} \label{thm:det}
    OP is deterministic, \emph{i.e.}, for any given battery level $e$ and channel state $h$, there exists an optimal action pair $(\rho^\star,\iota^\star)$ that is chosen with probability one to obtain the maximum reward.
\end{thm}

Formally, the previous theorem states that, using OP,~
\begin{align}
    f(\rho,\iota ; e,h) = \delta_{\rho,\rho^\star}\cdot \delta(\iota - \iota^\star),\qquad \forall e \in \mathcal{E}, \forall h \in \mathcal{H}
    \label{eq:f_mu_star}
\end{align}

\noindent where $\delta_{\cdot,\cdot}$ is the Kronecker delta function and $\delta(\cdot)$ is the Dirac delta function.
In the remainder of this section we prove Theorem~\ref{thm:det} in two steps. We first show that in the maximization process of Equation~\eqref{eq:mu_star} it is sufficient to focus on the second term $j_\mu(e)$ of Equation~\eqref{eq:G_mu}. This is a consequence of Proposition~\ref{propos:P_E_k}. Later, we apply a Lagrangian approach to show that a deterministic policy maximizes $j_\mu(e)$.

\begin{propos}\label{propos:P_E_k}
    $\mathbb{P}(E_k = e | \mathbf{S}_0)$ depends upon the policy only through $\mathbb{E}_{H}[f(\rho,\iota ; e,H)]$ for $\rho = 0,\ldots,e$, $\forall \iota \in [0,1]$ and $\forall e \in \mathcal{E}$.
\begin{proof}
The proof is by induction on $k$.
At $k = 0$, $\mathbb{P}(E_0 = e | \mathbf{S}_0 = (e_0,h_0))$ is equal to $1$ if $e = e_0$ and to $0$ otherwise. In this case there is no dependence upon the policy.

        Suppose that the thesis is true for $k$ (inductive hypothesis). Using the chain rule, the probability that $E_{k+1} = e'$ given the initial state is~
        \begin{align*}
            \mathbb{P}(E_{k+1} = e' | \mathbf{S}_0) = \sum_{e = 0}^{e_{\rm max}} \mathbb{P}(E_{k+1} = e' | E_k = e)\mathbb{P}(E_k = e | \mathbf{S}_0)
        \end{align*}
        
        Therefore, it is sufficient to show that $\mathbb{P}(E_{k+1} = e' | E_k = e)$ depends upon the policy only through the expectations $\mathbb{E}_H[f(\rho,\iota ; e,h)]$.
        Now, define the set $\mathcal{X}_{e',e} \triangleq \{(\rho,\iota,b) : \min\{y_T((1-\iota)b-\rho,e),e_{\rm max}\} = e' \}$. We have~
        \begin{align*}
            &\mathbb{P}(E_{k+1} = e' | E_k = e) \\ 
            &=\sum_{(\rho,\iota,b) \in \mathcal{X}_{e',e}} p_B(b) \sum_{h \in \mathcal{H}} p_H(h) f(\rho,\iota ; e,h) \\
            &=\sum_{(\rho,\iota,b) \in \mathcal{X}_{e',e}} p_B(b) \mathbb{E}_H[f(\rho,\iota ; e,H)]
        \end{align*}

        Thus, $\mathbb{P}(E_{k+1} = e' | E_0)$ depends upon the policy only through the expectations $\mathbb{E}_H[f(\rho,\iota ; e,h)]$.
\end{proof}
\end{propos}

For every $e \in \mathcal{E}$, $\rho = 0,\ldots,e$ and $\iota \in [0,1]$, fix a value $\alpha(\rho,\iota ; e)$. Consider now the set of policies $\Xi$ such that $\mathbb{E}_{H}[f(\rho, \iota ; e,H)]$ is equal to $\alpha(\rho, \iota ; e)$ for every $\mu \in \Xi$. 
Thanks to Proposition~\ref{propos:P_E_k}, in Equation~\eqref{eq:G_mu} $\pi_\mu(e)$, $\forall e \in \mathcal{E}$, is the same for every $\mu \in \Xi$. Thus, the maximization of $G_\mu$ can be rewritten as $\max_\mu \sum_{e \in \mathcal{E}} \pi_\mu(e) j_\mu(e) = \sum_{e \in \mathcal{E}} \pi_\mu(e) \max_\mu{j_\mu(e)}$, \emph{i.e.}, $G_\mu$ is maximized when $j_\mu(e)$ is maximized for every $e$. The maximization problem becomes a set of $e_{\rm max}+1$ simpler optimization problems. 

Focus now on a fixed $e \in \mathcal{E}$. We want to find the optimal $f(\rho,\iota ; e,h)$ that maximizes~
\begin{align*}
    j_\mu(e) = \mathbb{E}_{B,H}[\mathbb{E}_{\Theta_{e,H},\Phi_{e,H}}[g(\Theta_{e,H} + B \Phi_{e,H},H)]]
\end{align*}

\noindent subject to the constraints~
\begin{subequations}
\begin{alignat*}{2}
    &\mathbb{E}_{H}[f(\rho, \iota; e,H)] = \alpha(\rho, \iota ; e), \qquad &&\rho = 0,\ldots,e, \qquad \iota \in [0,1]
\end{alignat*}
\end{subequations}

We now want to show that the optimal $f(\rho,\iota ; e,h)$ has a deterministic structure using a Lagrangian approach. The Lagrangian function is~
\begin{align*}
    &\pushright{\mathcal{L}(e) = \mathbb{E}_{B,H} \Bigg[\sum_{\rho = 0}^{e}\left( \int_{0}^1 f(\rho,\iota ; e,H) g(\rho+B\iota,H) \mbox{d}\iota \right)}\Bigg]\\
    &\pushright{- \mathbb{E}_H \left[\sum_{\rho = 0}^{e} \int_{0}^{1} v_{\rm th}(\rho, \iota ; e) f(\rho,\iota ; e,H)\mbox{d}\iota \right]} \\
    &\pushright{= \mathbb{E}_{H} \Bigg[\sum_{\rho = 0}^{e}\left( \int_{0}^1 f(\rho,\iota ; e,H) \mathbb{E}_B\left[ g(\rho+B\iota,H)\right] \mbox{d}\iota \right)}\Bigg]\\
    &\pushright{- \mathbb{E}_H \left[\sum_{\rho = 0}^{e} \int_{0}^{1} v_{\rm th}(\rho, \iota ; e) f(\rho,\iota ; e,H)\mbox{d}\iota \right]}\\
    &\pushright{= \mathbb{E}_{H} \Bigg[\sum_{\rho = 0}^{e} \int_{0}^1 f(\rho,\iota ; e,H)\left(\tilde{g}(\rho,\iota,H) - v_{\rm th}(\rho,\iota ; e)\right) \mbox{d}\iota \Bigg]}
\end{align*}

\noindent where $v_{\rm th}(\rho,\iota ; e)$ is the Lagrange multiplier associated with the constraint $\mathbb{E}_{H}[f(\rho, \iota; e,H)] = \alpha(\rho, \iota ; e)$. Note that we neglected the terms $\alpha(\rho, \iota ; e)$ in the definition of $\mathcal{L}(e)$ because they do not contribute in the Lagrangian maximization process.
We also introduced a new function $\tilde{g}(\rho,i,h) \triangleq \mathbb{E}_B[g(\rho + B\iota,h)]$ that summarizes the contributions of the energy arrivals. Note that we assume not to know the realization of $B$ but we know its statistics (that is used to compute $\tilde{g}(\rho,\iota,h)$), thus OP depends upon the energy arrival statistics.

The structure of Equation~\eqref{eq:f_mu_star} is proved by taking, for a fixed $e$ and $h$, $(\rho^\star,\iota^\star) = \arg \max_{(\rho,\iota)_e}\Big(\tilde{g}(\rho,i,h) - v_{\rm th}(\rho,\iota ; e)\Big)$.\footnote{The idea behind this result is the following. Consider the maximization over $\xi\in [0,1]$ of $\xi g_1 + (1-\xi)g_2$, where $g_1$ and $g_2$ are two fixed numbers. The maximum is obtained when $\xi = 1$ if $g_1 > g_2$ and $\xi = 0$ otherwise. In our case $\xi$ is substituted by $f(\rho,\iota ; e,H)$ and $g_1$, $g_2$ by $\tilde{g}(\rho,i,h) - v_{\rm th}(\rho,\iota ; e)$.}

In this section we proved that for every state of the system, there exists an optimal pair that has to be always chosen to obtain the maximum long-term average reward. It is sufficient to find the optimal pair for every system state to define the optimal policy.

We now derive another property of OP, that is useful to perform the numerical evaluation.

\begin{propos} \label{propos:admissible}
    The optimal policy induces a Markov Chain with at most one recurrent class if $p_H(0) > 0$.
    \begin{proof}
        The Markov Chain has two dimensions: the battery and the channel. Since the channel gain is not controlled by the device and changes over time, it is always possible to move along this dimension. We now want to show that it is always possible to go from $e \in \{0,\ldots,e_{\rm max}-1\}$ to $e_{\rm max}$ and that the recurrent class is defined by the states that are reachable from $e_{\rm max}$.
        
        Since $p_H(0) > 0$, for some $k$, we have $H_k = 0$ and $E_k < e_{\rm max}$. In this case, the best policy is to choose $P_k = 0$ and $I_k = 0$, otherwise energy would be wasted without increasing the reward, which is sub-optimal. Thus, there exists a positive probability to go from state $(e,0)$ to $(e',h')$, with $e' \geq e+1$ (no transmissions are performed and new energy quanta arrive) and $h'$ a generic channel state. The reasoning can be iteratively applied to $e'$ until $e_{\rm max}$ is reached. We showed that it is always possible to increase the state along the battery dimension. The converse (decreasing the state along the battery dimension) in general is not true.
        
        In summary, we have a recurrent class composed of the last part of the MC in the battery dimension and, possibly, a transient class for low $e$.
    \end{proof}
\end{propos}

The proposition guarantees that the problem is well posed, \emph{i.e.}, it is independent of the initial state $\mathbf{S}_0$. In reality, the channel gain is a continuous quantity. In this case, we can assume that the device does not transmit for very low channel gains in order to guarantee a minimum SNR level.

\section{Numerical Evaluation} \label{sec:numerical_evaluation}

We use $g(x,H_k) = \ln(1+ \Lambda H_k x)$ as the reward function, thus $G_\mu$ represents the throughput of the system. To model the channel gain, we discretize an exponentially distributed variable (Rayleigh fading). We consider an energy arrivals process described by a truncated geometric r.v. of mean $\bar{b} = 20$ and maximum energy arrivals $b_{\rm max} = 50$. In this section we want to show the importance of applying the optimal transmission policy to obtain high performance. In particular, we show that the optimal policy computed without considering the inefficiencies of a real battery is strictly sub-optimal.

Several transmission techniques can be applied. Here, we compare three approaches: 1) Optimal Policy (OP) defined in Sec.~\ref{sec:optimization}, 2) optimal policy of an ideal battery (OP-IDEAL) and 3) Greedy Policy (GP). OP-IDEAL is the optimal policy derived considering an ideal battery (without storage losses) and then applied to a real battery (with storage losses). GP is a policy that always poses $I_k = 1$ (the harvested energy is directly sent into the channel and the presence of the battery is neglected). This choice could be interesting for very inefficient or small batteries, but, in general, provides low rewards.

\begin{figure}[t]
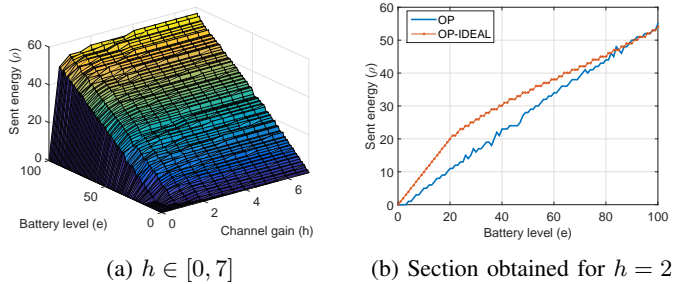

    \centering
    \begin{subfigure}{0.5\columnwidth}
        \includegraphics[trim = 0mm 0mm 1mm 6mm,  clip, width=1\columnwidth]{Sent_Energy_e_h.eps}~
        \caption{$h \in [0,7]$} \label{fig:Sent_Energy_e_h}
    \end{subfigure}~
    \hspace*{\fill}
    \begin{subfigure}{0.5\columnwidth}
        \includegraphics[trim = 1mm 0mm 0mm 6mm,  clip, width=1\columnwidth]{Sent_Energy_e.eps}
        \caption{Section obtained for $h = 2$} \label{fig:Sent_Energy_e}
    \end{subfigure} \\[10pt]
    \caption{Sent energy $\rho(e,h)$ of OP as a function of the battery state and of the channel gain.}
    \label{fig:Sent_Energy}
\end{figure}

Initially, suppose $I_k = 0$, \emph{i.e.}, the harvested energy cannot be sent directly. 
In \figurename~\ref{fig:Sent_Energy_e_h}, we show the amount of sent energy $\rho$ of OP for every state of the system when $e_{\rm max} = 100$, $\Lambda = 0.1$ and $\beta_{\rm n.l.} = 1.05$. It can be seen that at $h = 0$ we have $\rho = 0$, as explained in Proposition~\ref{propos:admissible}. An interesting point is that the sent energy increases fast when the channel gain is very low and later becomes almost channel independent. This fact can be used to design a low-complexity policy. 

\begin{figure}[t]
  \centering
  \includegraphics[trim = 0mm 0mm 0mm 3mm,  clip, width=0.8\columnwidth]{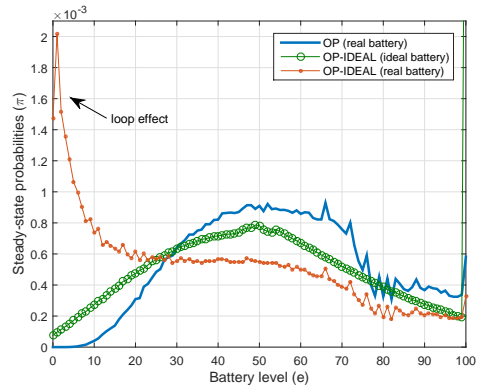}
  \caption{Steady-state probabilities $\pi(e,h)$ of OP and OP-IDEAL (for an ideal as well as a real battery) as a function of the battery state when $h = 2$.}
  \label{fig:pi_e}
\end{figure}

In \figurename~\ref{fig:Sent_Energy_e} we present a section of the sent energy, obtained when $h = 2$. Here we also represent OP-IDEAL in order to compare the two policies. Note that, when $e$ is not very high, the OP curve is lower than the other. In practice, OP tends to transmit with low powers in order to avoid the low energy states (that have high storage losses). This results (see \figurename~\ref{fig:pi_e}) in $\pi_\mu(e) = 0$ when $e$ is very low (as stated in Proposition~\ref{propos:admissible}, there is only one recurrent class). 
This is particularly important because, due to the energy losses, a \emph{loop effect} may happen. We remember that when $e$ is low or high, a lot of energy is wasted in the recharging process (see the structure of $s(y)$). In particular, focus on the low energy states. If a high power transmission is performed in these states (as per OP-IDEAL), the battery does not manage to be recharged (the transmissions almost compensate the energy arrivals). In practice, the SOC is trapped in the low energy states.
Note that also for the high energy states the energy losses are high, but in this case the loop effect is not present because it is always possible to empty the battery, thus moving the energy state to a more advantageous region. These behaviors can be seen in the steady-state probabilities depicted in \figurename~\ref{fig:pi_e}. When we apply OP-IDEAL to the real battery, we can notice the loop effect for the low energy states. This is particularly critical because it degrades the system performance (if $e$ is low, only little energy can be used for transmission, resulting in low rewards) and the device availability (for most of the time EHD is almost in outage). 
The circled green line represents the steady-state probabilities for an ideal battery. This curve is smoother than the others because the energy arrival distribution is independent of $e$, whereas in the other cases, for every energy state we have a different energy arrival distribution according to Equation~\eqref{eq:s_x_y}. In an ideal battery, the probability of being in the final energy state is high because a lot of energy arrives with respect to the battery size. By increasing the size, the peak decreases. In a real battery instead, since the storage losses are high for high $e$, the peaks are much smaller.

OP-IDEAL and GP have long-term average throughput equal to $84\%$ and $78\%$ of the throughput of OP. In this case, using OP-IDEAL for a real battery is not convenient, since it has almost the same throughput of GP, that is much simpler to compute and easier to implement.

\begin{figure}[t]
  \centering
  \includegraphics[trim = 0mm 0mm 0mm 6mm,  clip, width=0.8\columnwidth]{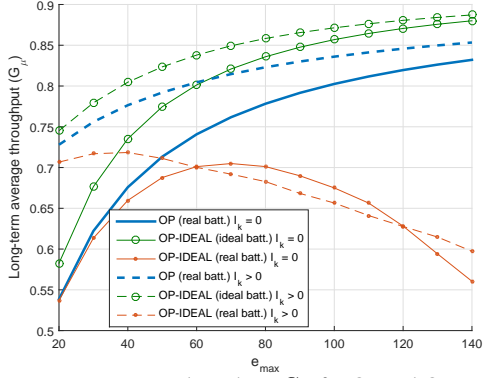}
  \caption{Long-term average throughput $G_\mu$ for OP and OP-IDEAL (for an ideal as well as a real battery) as a  function of the battery size $e_{\rm max}$.}
  \label{fig:throughput}
\end{figure}

\figurename~\ref{fig:throughput} shows the effects of a finite battery size on the throughput. Focus on the continuous lines. We plot the rewards for OP and OP-IDEAL (for an ideal as well as a real battery). As expected, the OP curve falls between OP-IDEAL in the real and ideal case. When the battery is very small with respect to the energy arrivals, OP-IDEAL and OP are close. This is because a lot of energy arrives and the battery is almost always full. Instead, as $e_{\rm max}$ increases, the gap between the two policies increases significantly. This is due to the loop effect because the larger the battery, the larger the region where the energy losses are high.
Consider the OP curve. It can be seen that it increases slowly for high $e_{\rm max}$, \emph{e.g.}, at $e_{\rm max} = 100$ we have a reward of $0.80$ and to increase this value to $0.83$ it is necessary to have a much larger battery ($e_{\rm max} = 140$). This happens because the effects of overflow and outage are reduced with large batteries.
Note that the reward saturates and does not grow unbounded.

We now want to show the improvements that can be obtained when $I_k > 0$. Intuitively, when the battery is charged in a high losses region, it is better to directly send the harvested energy into the channel. However, note that, as seen in the previous example, OP already avoids the inefficient regions (the low SOC states are rarely reached).

In \figurename~\ref{fig:throughput}, with dashed lines, we show the system throughput when $I_k$ is not forced to zero. For OP and OP-IDEAL for an ideal battery, as expected, the curves are above their counterparts with $I_k = 0$. Instead, for OP-IDEAL for a real battery, it may be better not to use $I_k > 0$ in some regions. Thus, as can be seen, even with $I_k > 0$, OP-IDEAL is strictly sub-optimal when applied to a real battery. 
Intuitively, allowing to directly send the harvested energy does not provide a great improvement when the battery is large: in this case it is sufficient to operate in a region with low energy losses in order to efficiently store the harvested energy and use it later. This is not possible in a small battery, and in this case directly sending the harvested energy may provide a significant reward improvement (left part of \figurename~\ref{fig:throughput}).

\begin{figure}[t]
  \centering
  \includegraphics[trim = 0mm 0mm 0mm 6mm,  clip, width=0.8\columnwidth]{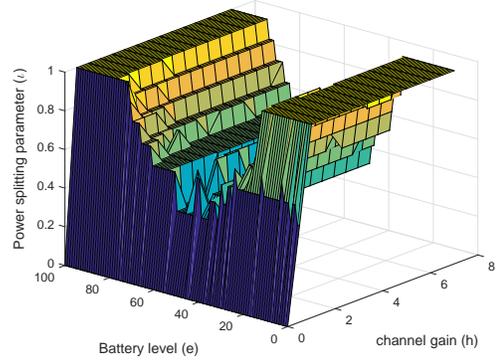}
  \caption{Power splitting parameter $\iota(e,h)$ for OP as a function of the battery state $e$ and of the channel state $h$.}
  \label{fig:i_e_h}
\end{figure}

With the parameters of \figurename~\ref{fig:Sent_Energy_e_h}, when we find OP with $I_k > 0$, the reward becomes $0.83$ (instead of $0.80$). Even if the relative improvement is small, note that, with $I_k = 0$, it is necessary to have a much larger battery to obtain a reward of $0.83$ ($e_{\rm max} = 140$). Thus, allowing $I_k > 0$ permits to obtain higher rewards without increasing the battery size.
In \figurename~\ref{fig:i_e_h} we show the power splitting parameter $\iota$ for every battery state and channel gain. When SOC is high or low, the energy losses are high, thus in these cases it is better to directly send the harvested energy. As $e$ moves toward $e_{\rm max}/2$, also $\iota(e,h)$ decreases until the mid region where $\iota(e,h) \approx 0.5$. In this case it is convenient to store the energy for future use because the losses are very low and, at the same time, to directly send energy into the channel to obtain an immediate reward. When the SOC is low, the battery can still be charged by exploiting the states where the channel gain is zero.

\section{Conclusions} \label{sec:conclusions}

We studied an Energy Harvesting Device with imperfect energy storage capabilities. During the energy saving process, part of the incoming energy is wasted and this effect depends upon the current state of charge of the device. We proved that, for given battery status and channel gain, the Optimal Policy is deterministic. We also showed that the underlying MC induced by OP has at most one recurrent class. These results were used to develop the numerical evaluation, where we studied how OP influences the performance of an EHD with a capacitor. We found that the energy losses degrade the system performance, making the device fall into a region where the loop effect may be present. Finding the optimal policy in this context is useful in order to avoid the loop effect and to restore high rewards. We noticed that the reward increases slowly with the battery size and is upper bounded. Finally, we showed that allowing $I_k > 0$ is equivalent, in terms of reward, to using a much larger battery, thus implementing the power splitting mechanism is convenient when the battery is small.

Possible extensions consist in the relaxation of some hypothesis, \emph{e.g.}, those about perfect SOC and channel knowledge. Also, the circuitry costs can be taken into account. Moreover, it would be interesting to study the effects of imperfect energy storage with other functions (in addition to~\eqref{eq:s_x_y}).

\bibliography{D:/OneDrive/Ph.D/Articoli/EHD}{}
\bibliographystyle{IEEEtran}

\end{document}